\documentclass[twocolumn,prd,amsmath,superscriptaddress,nofootinbib,floatfix]{revtex4}

\usepackage{amsmath}
\usepackage{bm}
\usepackage{graphicx}
\usepackage{color}
\usepackage{subfigure}

\newcommand{\ave}[1]{\left\langle#1\right\rangle}

\def\gsim{\;\rlap{\lower 2.5pt
 \hbox{$\sim$}}\raise 1.5pt\hbox{$>$}\;}
\def\lsim{\;\rlap{\lower 2.5pt
 \hbox{$\sim$}}\raise 1.5pt\hbox{$<$}\;}

\def\be{\begin{equation}}
\def\ee{\end{equation}}
\def\bea{\begin{eqnarray}}
\def\eea{\end{eqnarray}}

\begin{document}

\title{Cross-correlating the Thermal Sunyaev-Zel'dovich Effect and the Distribution of Galaxy Clusters}

\author{Wenjuan Fang}
\affiliation{Department of Physics, University of Michigan, Ann Arbor, MI 48109}
\author{Kenji Kadota}
\affiliation{Department of Physics, University of Michigan, Ann Arbor, MI 48109}
\affiliation{Department of Physics, Nagoya University, Nagoya 464-8602, Japan}
\author{Masahiro Takada}
\affiliation{Institute for the Physics and Mathematics of the Universe (IPMU),
The University of Tokyo, Chiba 277-8582, Japan}

\date{\today}

\begin{abstract}
We present the analytical formulas, derived based on the halo model, 
to compute
the
cross-correlation between the thermal Sunyaev-Zel'dovich (SZ) effect and the
distribution of galaxy clusters. By binning the clusters according to
their redshifts and masses, this cross-correlation, the so-called stacked
SZ signal, reveals the average SZ profile around 
the 
clusters. The stacked SZ signal 
is obtainable from 
a joint analysis of an
arcminute-resolution
 cosmic microwave background (CMB) experiment 
and 
an overlapping optical survey,
which allows for
detection of 
the SZ signals for clusters whose masses are below the individual-cluster detection
 threshold.
We 
derive the error covariance matrix for measuring the stacked SZ signal, and 
then
forecast for
its detection from ongoing and forthcoming combined CMB-optical surveys. We find that, over a
wide range of mass and redshift, the stacked SZ signal can be detected
with a significant signal to noise ratio (total $S/N\gsim 10$), whose value peaks for the clusters with intermediate masses and redshifts. Our calculation also shows that the stacking
method allows for probing the clusters' SZ profiles over a wide range of scales,
even out to projected radii as large as the virial radius,
thereby providing a promising way to study gas physics at the outskirts
of galaxy clusters.
\end{abstract}

\maketitle

\section{Introduction}

Galaxy clusters are the most massive, gravitationally bound objects in
the universe, and therefore, their properties, e.g., the cluster abundance and its redshift evolution, are very
sensitive to the underlying cosmology including the nature of dark energy
\cite{Vikhlininetal:09,Rozoetal:10}. The Sunyaev-Zel'dovich (SZ) effect
\cite{SZ72} offers a unique means of finding clusters out to high
redshift due to its redshift independence \citep[see][for a
review]{Car02}. The wide-field, arcminute-resolution experiments of the 
cosmic microwave background (CMB) such as the Planck satellite
\cite{PlanckSZ:11}, the Atacama Cosmology Telescope (ACT)
\cite{Fowleretal:10}, and the South Pole Telescope (SPT) \cite{SPT:10},
have been drawing a great attention, because such CMB surveys 
enable us 
to construct a catalog of SZ clusters under a homogeneous selection
function out to high redshifts beyond $z=1$. Such a cluster catalog can
in turn be used to carry out the high-precision cosmological probes \citep[][also see
references therein]{Haimanetal:01,LimaHu:05,OguTak11}.

In the light of cluster cosmology, combining the SZ survey with
a multi-color optical survey offers a great synergy, if the two
cover the same region of the sky. First, the multi-color optical
survey 
provides an 
estimate for the redshift of each SZ-discovered cluster 
from the photometric redshifts of its member galaxies. Otherwise
the cluster's redshift is unknown due to the redshift independence of the SZ
signal.  Second, from such an optical survey, we can construct a
catalog of clusters by finding the concentration of galaxies in angular
and photometric redshift space 
\cite{GladdersYee:00,Koesteretal:07}. 
Hence, by combining the SZ and optical surveys,
 we can 
obtain 
a more robust
catalog of clusters. Third,
 from 
the joint experiment, 
we can 
explore the cluster mass-observable scaling relation by
combining various observables including the measured SZ signal strength
and the optical richness (i.e. the number of member galaxies). 
We  
can also include the weak lensing measurement on an individual cluster basis
\cite{Okabeetal:10b,Marroneetal:11}, where the lensing signal is
available from the same optical data set,  by
exploring in detail the lensing shearing effects on background galaxy
shapes due to the foreground cluster  \citep[see, e.g.,][for a weak
lensing study of galaxy clusters]{Okabeetal:10}.  The use of a well-calibrated
mass-observable relation is of critical importance to attain the
high-precision of cluster cosmology~\cite{LimaHu:05,DETF}

By cross-correlating the SZ map with the distribution of
optically selected clusters for a sufficiently large cluster sample,
one can {\em statistically} measure the SZ signals of these clusters -- the so-called {\em
stacked} SZ signal. The stacking method 
enables us to measure
the SZ
signals for halos down to relatively low masses as well as up to large projected radii,
which otherwise are very difficult to measure on an individual cluster basis,
because the signals are too small. In particular, the stacking method
allows a {\em tomographic} reconstruction of the SZ signal as a
function of redshift~\cite{pen2,shao1}, while the CMB signal itself is by nature
two-dimensional, reflecting the information projected between the last
scattering surface and the observer. The stacked SZ signals are useful for 
studies of the {\em mean} cluster mass-observable relation thanks to the
statistically-boosted signal-to-noise ratios. Furthermore, the stacked
SZ signals can be used to study the intracluster gas physics and to falsify and/or refine our theoretical understanding of galaxy clusters
\cite{li2,Shawetal:10,Lau09,FanHai07,NagKV07,Bat10}.

The initial attempts to measure the stacked SZ signals have been
recently reported, including those by combining the WMAP or Planck data with the SDSS maxBCG cluster 
catalog \cite{scott1,pl3} and the ACT data with the SDSS luminous
red galaxies (LRGs) \cite{Handetal:11}, which preferentially reside in massive halos. Such measurements will be further significantly improved by combining the arcminute-resolution,
high-sensitivity CMB surveys with overlapping wide-field, deep optical surveys, e.g.,
the SPT survey with the Dark Energy Survey (DES) \cite{DES}, and the ACT with
the Subaru Hyper Suprime-Cam (HSC) Survey \cite{Miyazakietal:06,OguTak11}\footnote{also see http://sumire.ipmu.jp/en/}.

Hence the purpose of this paper is to study the prospects of measuring
the stacked SZ signals from ongoing and upcoming combined SZ and optical
surveys. To make predictions for the stacked SZ profiles, we 
model the distribution of the hot gas contributing to the SZ effect
\cite{KomSel02} as well as the 
 distribution of galaxy
clusters according to the halo model \cite{CooShe02}. Our formulas should be useful for future stacked SZ analysis. 
Following \cite{OguTak11}, we derive the
covariances of the stacked SZ measurements for clusters binned by their redshifts and masses, which quantify the
statistical 
uncertainties in the measurements.

The outline of the paper is as follows. In Sec. \ref{sec:calculation}, we present our derivation for the formulas to calculate the cross-correlation between the SZ effect and the distribution of galaxy clusters, and give the covariance matrix for its measurement. In Sec. \ref{sec:detection}, we forecast the prospects of measuring the stacked SZ signals from several ongoing and upcoming surveys, and discuss our results. Finally, we conclude in Sec. \ref{sec:discuss}. Throughout this paper, we adopt the best-fit flat
$\Lambda$CDM model from WMAP 7-yr results \cite{WMAP7} with
$\Omega_m=0.27$, $\Omega_b=0.045$, $h=0.71$, $\sigma_8=0.8$, $n_s=0.96$.

\section{Theoretical Calculation}
\label{sec:calculation}

In this section, following the method developed in
\cite{OguTak11} and \cite{ColKai88} 
\citep[also see][]{KomKit99,SchBer91,PeacockSmith:00,Seljak:00,MaFry:00,Sco01,CooShe02,TakadaBridle:07}, we use the halo model to calculate the cross-correlation between the SZ effect and the distribution of galaxy clusters, and give the covariance matrix for its measurement.

\subsection{The SZ Effect and the SZ Power Spectrum}
\label{sec:szeffect}

The thermal SZ effect introduces the following fractional distortion to the CMB
temperature measured at frequency $\nu$ and in the angular direction of $\hat{n}$~\citep[see e.g.][]{SZ72,Rep95,Car02}:
\begin{equation}
\frac{\Delta T}{T}(x,\hat{n})=f(x)y(\hat{n}),
\end{equation}
where we have defined the dimensionless frequency $x\equiv h\nu/k_B T$, with $h$ the Planck constant, $k_B$ the Boltzmann constant, and the frequency dependence of the effect $f(x)$ is given by
\begin{equation}
f(x)=x\coth\left(\frac{x}{2}\right)-4,
\end{equation}
where relativistic corrections \citep[see e.g.][]{RepII95,ChaLas97,SazSun98,Ito98} have been neglected. Note, in the Rayleigh-Jeans (RJ) limit, $f(x)\rightarrow -2$. The Compton $y$-parameter $y(\hat{n})$ is proportional to the electron pressure $P_e$ projected along the line of sight: 
\begin{equation}
y(\hat{n})=\frac{\sigma_T}{m_e c^2}\int_{\hat{n}} P_e d\ell,
\end{equation}
where $\sigma_T$ is the Thomson cross section, $m_e$ is the electron mass, and $c$ is the speed of light.

Using the halo model, $y(\hat{n})$ can be expressed as
\begin{eqnarray}
y(\hat{n})&=&\frac{\sigma_T}{m_e c^2}\int \frac{c
 dz}{H(z)}\frac{1}{(1+z)}\int d M_h \int d^3 x_h \nonumber \\
&&\hspace{-2em}\times
\sum_i
\delta_D(M_h-M_i)\delta_D^3(\vec{x}_h-\vec{x}_i)P_e(|\vec{x}-\vec{x}_h|,M_h,
z_h),\nonumber\\
\label{eqn:y}
\end{eqnarray}
where $H(z)$ is the Hubble expansion rate at redshift $z$; $\delta_D$
stands for the Dirac delta function;
$M_i$ and $\vec{x}_i$ are the mass and center position of the $i$-th halo
on the light cone, and the sum is done over all the haloes;
 $P_e(|\vec{x}-\vec{x}_h|,M_h, z_h)$ is the pressure at position
$\vec{x}$
of the electrons in the dark matter halo with mass $M_h$, redshift $z_h$, and centered at
$\vec{x}_h$.
Note that the 3-dimensional position $\vec{x}$ is specified by 
$(\chi, \hat{n})$, where $\chi$ is the comoving radial
distance corresponding to redshift $z$ on the light cone.

By using 
the flat-sky approximation and the Limber approximation \cite{Limber:54}, 
we obtain the SZ power spectrum \cite{ColKai88,KomKit99}:
%
\begin{equation}
C_{\ell}^{\rm yy}=C_{\ell}^{\rm yy,1h}+C_{\ell}^{\rm yy,2h},
\end{equation}
with the one-
and two-halo term contributions,
 $C_{\ell}^{\rm yy,1h}$ and $C_{\ell}^{\rm yy,2h}$, given by
\begin{eqnarray}
C_{\ell}^{\rm yy,1h}&=&f(x)^2\int dz 
\frac{d^2V}{dzd\Omega}\int dM \frac{dn}{dM}(M,z)|\tilde{y}_{\ell}(M,z)|^2,\nonumber\\
C_{\ell}^{\rm yy,2h}&=&f(x)^2\int dz \frac{d^2V}{dzd\Omega}P_m^L\left(k=\frac{\ell}{\chi},z\right)W^y_{\ell}(z)^2,
\label{eq:clsz}
\end{eqnarray}
where $d^2V/dzd\Omega=c\chi^2/H$ is the comoving volume element, 
$dn/dM$ is the halo mass function, $\tilde{y}_{\ell}(M,z)$ is the 2D
Fourier transform of the Compton $y$-parameter profile of a cluster with
mass $M$ and at redshift $z$, 
and $P_m^L(k,z)$ is the linear matter power spectrum.
The function $W^y_{\ell}(z)$ is defined as
\begin{equation}
W^y_{\ell}(z)\equiv \int dM \frac{dn}{dM}(M,z) b(M,z) \tilde{y}_{\ell}(M,z),
\end{equation}
where $b(M,z)$ is the halo bias. As can be seen from Eq.~(\ref{eq:clsz}), the SZ power spectrum arises from the contributions of halos
over all the redshift range, and therefore contains only the projected information.

Below in our calculation, we would use the halo mass function from \cite{Jen01},
specifically their Eq.~(B3), and the halo bias from
\cite{SheTor99,HuKra03}. To calculate the linear matter power spectrum,
we use the transfer function from \cite{EH99}. Our model for the
intra-cluster gas is taken from \cite{KomSel02}, where the gas is
assumed to be in hydrostatic equilibrium with the dark matter potential,
have a constant polytropic equation of state, and trace the dark matter
density with a constant amplitude offset in the outer parts of a
cluster. 
We use the virial overdensity \cite{BryNor98,Kuh05} to define the cluster mass, and 
truncate 
the gas profile 
at
three times the virial radius $r_{\rm vir}$. As implied from hydrodynamic simulations, the gas profile may extend beyond $r_{\rm vir}$ due
to various nonlinear processes such as shock heating, feedback from star
formation etc. \cite{BryNor98, Kravtsovetal:05}. Hence, it would be
interesting to study the gas properties out to these large radii,
which can be explored by using the stacking analysis (see below for further
discussion).

\subsection{The Distribution of Galaxy Clusters and Its Power Spectrum}

Given a cluster survey, the galaxy clusters can be binned according to
their observed redshift $z_{\rm obs}$ and estimated mass 
$M_{\rm obs}$. Given $P(z_{\rm obs}|M,z)$ and $P(M_{\rm obs}|M,z)$, i.e. the
probability distribution functions (PDFs) of observing $z_{\rm obs}$ and
$M_{\rm obs}$ for a cluster with true redshift $z$ and true mass $M$,
the probability that the cluster is selected to the ``$a$''-th redshift
bin with 
$z_{\rm obs}\in[z^a_{\rm obs,min},z^a_{\rm obs,max}]$ and the ``$b$''-th mass bin with $M_{\rm obs}\in[M^b_{\rm obs,min},M^b_{\rm obs,max}]$ can be calculated by
\begin{eqnarray}
S_{ab}(M,z)&=&\int_{z_{\rm obs, min}^a}^{z_{\rm obs, max}^a}dz_{\rm
 obs}P(z_{\rm obs}|M,z)\nonumber\\
&&
\times \int_{M_{\rm obs, min}^b}^{M_{\rm obs, max}^b}dM_{\rm obs}P(M_{\rm obs}|M,z),
\end{eqnarray}
where $S_{ab}$ is called the selection function. In the ideal case that both $z$ and $M$ can be measured accurately,
which we would assume in the following calculation for simplicity, the PDFs become delta functions, and $S_{ab}$ reduces to 
\begin{eqnarray}
S_{ab}(M,z)&=&\Theta(z-z^a_{\rm obs,min})\Theta(z^a_{\rm obs,max}-z)
\nonumber\\&&\hspace{-2em}
\times
\Theta(M-M^b_{\rm obs,min})\Theta(M^b_{\rm obs,max}-M),
\end{eqnarray}
where $\Theta$ stands for the Heaviside step function.

With the halo model formulation, the 2-dimensional angular number density 
of galaxy clusters in the
``$(ab)$''-th bin can be calculated by
\begin{eqnarray}
n_{ab}^{2D}(\hat{n})&=&\int dz \frac{d^2V}{dzd\Omega} \int dM
 S_{ab}(M,z) \nonumber\\&&
\times \sum_i \delta_D(M-M_i)\delta_D^3(\vec{x}-\vec{x}_i)\label{eqn:n2D},
\end{eqnarray}
where all symbols have the same meaning as before.
Its ensemble average is given by
\begin{equation}
\bar{n}_{ab}^{2D}=\int dz \frac{d^2V}{dzd\Omega} \int dM
 S_{ab}(M,z)\frac{dn}{dM}(M,z),\label{eqn:nbar2D}
\end{equation}
where we have used $\ave{ \sum_i
\delta_D(M-M_i)\delta_D^3(\vec{x}-\vec{x}_i)}=dn/dM$ \cite{Sco01}.  

With similar approximations as before, 
the power spectrum for 
the distribution of two cluster samples that are in the
``$(ab)$''-th and ``$(a'b')$''-th bins, respectively, can be calculated by
%
\begin{equation}
C_{\ell,(ab,a'b')}^{\rm hh}=\int dz \frac{d^2V}{dzd\Omega}P_m^L\left(k=\frac{\ell}{\chi},z\right)W^h_{ab}(z)W^h_{a'b'}(z),
\end{equation}
where $W^h_{ab}(z)$ is defined as 
\begin{equation}
W^h_{ab}(z)\equiv \frac{1}{\bar{n}^{2D}_{ab}}\int dM \frac{dn}{dM}(M,z)S_{ab}(M,z)b(M,z).
\end{equation}
As long as the redshift of clusters can be measured accurately, as
we assume in this paper, only 
the clusters in the same redshift
bin would correlate with one another, so we have 
$C_{\ell,(ab,a'b')}^{\rm hh}=\delta_{aa'}^KC_{\ell,(ab,ab')}^{\rm
hh}$, where $\delta^K_{aa'}$ is the Kronecker delta function.
Note in this case, $W^h_{ab}(z)=0$ if $z$ is outside the ``$a$''-th redshift bin. Hence, with redshift information,
the cluster power spectrum is free of projection effect.

\subsection{The Cross-Correlation: the Stacked SZ Profile}
\label{subsec:clyh}

With 
Eqs.~(\ref{eqn:y}) and (\ref{eqn:n2D})
 in hand, we are now ready to calculate the cross-correlation between
 the SZ signal (to be explicit, $\Delta T^{\rm SZ}/T_{\rm CMB}$) and the
 cluster distribution. 
Again by adopting the flat-sky approximation and  
the Limber approximation, 
the cross power spectrum for the SZ signal and the distribution of
galaxy clusters in the ``($ab$)''-th bin can be calculated by \cite[also see][for the method of derivation]{OguTak11}
\begin{equation}
C_{\ell,(ab)}^{y{\rm h}}=C_{\ell,(ab)}^{y{\rm h,1h}}+C_{\ell,(ab)}^{y{\rm h,2h}},
\end{equation}
where the one- and two-halo term contributions are given by
\begin{eqnarray}
C_{\ell,(ab)}^{y{\rm h,1h}}&=&\frac{f(x)}{\bar{n}^{2D}_{ab}}\int dz
 \frac{d^2V}{dzd\Omega}\int dM \frac{dn}{dM}(M,z)\nonumber\\
&&\times S_{ab}(M,z)\tilde{y}_{\ell}(M,z),\label{eqn:clyh1h}
\end{eqnarray}
\begin{equation}
C_{\ell,(ab)}^{y{\rm h,2h}}=f(x) \int dz \frac{d^2V}{dzd\Omega}P_m^L\left(k=\frac{\ell}{\chi},z\right)W_{ab}^h(z)W^y_{\ell}(z).
\end{equation}
This is the Fourier-transform of the so-called stacked SZ profile,
which practically, can be obtained from observation by averaging
the SZ maps around the cluster centers and then averaging over the orientation. Hence, the center of cluster should be known {\em a priori}, e.g., from the position of the
brightest cluster galaxy in the case that the cluster catalog is
constructed from an optical survey as we have had in mind. Offsets of the ``cluster centers''
would cause 
dilution of the stacked signal at small-angles (high-multipoles). 
Since the
effect can be straightforwardly included by using the method developed
in \cite{OguTak11}, we here ignore it for
simplicity. 
We also notice that real clusters usually have aspherical shapes and substructures, which are not taken into account in our halo model calculation. A careful study of their effects on the stacked SZ profile would involve comparison with hydrodynamical simulations (see, e.g., \cite{Bat11} for their effects on the SZ power spectrum), which are beyond the scope of the current paper. 

Comparing with Eq.~(\ref{eq:clsz}), one can find that, since the
selection function $S_{ab}(z)$ or $W^{\rm h}_{ab}(z)$ vanishes outside the
redshift bin of the clusters taken for the cross-correlation, the
cross-power spectrum contains only the SZ contributions from clusters in the
redshift bin. Put another way, the cross-correlation method allows a
tomographic reconstruction of the SZ signal as a function of
redshift \citep[also see][for similar discussions]{shao1,pen2}.
When the redshift slice is sufficiently narrow, the cross-power
spectrum is equivalent to the three-dimensional power spectrum through
the correspondence of $k=\ell/\chi$, under the assumption of statistical isotropy.

Perhaps more intuitive is the Fourier-transformed counterpart of the
cross-power spectrum, i.e., the stacked y-profile, defined as
\begin{equation}
\ave{y}_{(ab)}\!(\theta) \equiv \frac{1}{f(x)}
\int \frac{\ell d\ell}{2\pi}J_0(\ell\theta)C_{\ell,(ab)}^{y{\rm h}},\label{eqn:xiyh}
\end{equation}
where $J_0(\ell \theta)$ is the zeroth order Bessel function. The stacked y-profile gives the average y-profile around the clusters in the bin, which have 
similar redshift
and similar mass, hence gives
the average projected pressure profile:
$\ave{\int\!dl~P_e(r,l)}\!(r_{\perp})=m_ec^2\ave{y}(\theta) /\sigma_T$, where
$r_{\perp}=\chi(z)\theta/(1+z)$. Under the assumption of statistical isotropy, 
the average 3-dimensional pressure profile is considered to be spherically symmetric, and therefore
can be recovered from the stacked y-profile through a deprojection
process. This is analogous to 
the average mass profile of the clusters obtained from 
the stacked lensing measurement
\cite{Johnstonetal:07,Okabeetal:10}.

\subsection{Measurement Uncertainties: the Covariance Matrix}
\label{subsec:covariance}

Given a survey, the stacked SZ signal can be measured by constructing an estimator from the observed fields. Uncertainties on the estimated values can be obtained from their covariance matrix. Similar to the covariance matrix for the stacked lensing
power spectrum 
\cite{OguTak11}~\citep[also see][]{Kno95,Sco99},
the covariance matrix for the stacked SZ power
 spectrum is given by
\begin{eqnarray}
{\rm Cov}\left(C_{\ell,(ab)}^{y{\rm h}},C_{\ell',(a'b')}^{
	  y{\rm h}}\right)&=&\frac{\delta^K_{\ell \ell'}}{f_{\rm
sky}(2\ell+1)\Delta\ell}\nonumber\\
&&\hspace{-8em}\times
 \left[\hat{C}_{\ell}^{yy}\hat{C}_{\ell,(ab,a'b')}^{\rm
  hh}+\hat{C}_{\ell,(ab)}^{y{\rm h}}\hat{C}_{\ell,(a'b')}^{y{\rm h}}\right],
\label{eq:cly_cov}
\end{eqnarray}
where $f_{\rm sky}=\Omega_s/4\pi$ is the fraction of the survey sky, with $\Omega_s$ the survey area,
$\Delta\ell$ is the width of the 
 $\ell$ band used to estimate the power spectrum, and $C_{\ell}$ 
with a hat symbol
denotes the corresponding power spectrum for the observed fields, i.e. 
including 
 noise contamination. 
Note, for simplicity, we have assumed the fields to be Gaussian in this
 calculation. See \cite{KomSel02,ZhangSheth:07} for discussions on the importance of non-Gaussian errors.

For a CMB experiment, the fluctuations in the temperature field that is actually
observed have contributions from the primary CMB anisotropy, as well as from various
secondary anisotropies and 
foregrounds 
such as 
 the thermal SZ effect, the kinetic SZ effect, radio point sources,
 infrared point sources, diffuse Galactic foregrounds etc, in addition
 to the instrumental noise \cite{TegEis00}. Since the thermal SZ effect has a different frequency-dependence from the other components, by combining the data from multiple frequencies, the thermal SZ signal can be extracted \cite[see e.g.][]{CooHu00}. 

Here, for definiteness, we adopt the minimum-variance subtraction technique presented in \cite{CooHu00} to extract the SZ signal, and, for simplicity, we include only the instrumental noise. Therefore, the observed SZ power spectrum is given by
\begin{equation}
\hat{C}_{\ell}^{yy}=C_{\ell}^{yy}+\frac{1}{\sum_c w_c s_c^2 B_{c\ell}^2},
\end{equation}   
where we use the subscript ``$c$'' to label the quantities for different frequency channels; $w=(\sigma_{T}\theta_{\rm FWHM}/T_{\rm CMB})^{-2}$, with $\sigma_T$ the rms of the instrumental noise per pixel, and $\theta_{\rm FWHM}$ the full width, half-maximum (FWHM) of the beam, which is assumed to be Gaussian; $B_{\ell}$ is the Fourier transform of the beam profile, and $B_{\ell}^2=\exp[-\ell(\ell+1)\theta_{\rm FWHM}^2/(8\ln{2})]$. Note we aim at extracting the SZ signal in the RJ limit, so the observation at each frequency has been rescaled by the ratio of frequency dependence: $s=-f(x)/2$. Specifications of the CMB experiments considered in this paper are given in Table~\ref{tab:cmb}. Since the instrumental noise is not correlated with the distribution of clusters, we have 
\begin{equation}
\hat{C}_{\ell}^{y{\rm h}}=C_{\ell}^{y{\rm h}}.  
\end{equation}
Finally, we include a shot noise in the observed cluster power spectrum:
\begin{equation}
\hat{C}_{\ell,(ab,a'b')}^{\rm hh}=C_{\ell,(ab,a'b')}^{\rm hh}+\frac{1}{\bar{n}^{2D}_{ab}}\delta^K_{aa'}\delta^K_{bb'}.
\end{equation}

Likewise, we can derive the covariance matrix for the stacked $y$-profile, $\ave{y}_{(ab)}\!(\theta_i)$, which is estimated by Fourier transforming the stacked SZ power spectrum and then averaging over an annular area around $\theta_i$ \citep[see e.g.][for a similar derivation of the covariance for the correlation functions of the lensing field]{TakadaJain:09}. 
The covariance between $\ave{y}_{(ab)}\!(\theta_i)$ and 
$\ave{y}_{(a'b')}\!(\theta_j)$ is found to be
\begin{eqnarray}
{\rm Cov}\left(
\ave{y}_{(ab)}\!(\theta_i),\ave{y}_{(a'b')}\!(\theta_j)
\right)
&=&\frac{1}{\Omega_sf(x)^2}\int\frac{\ell
d\ell}{2\pi}\bar{J}_0(\ell\theta_i)
\nonumber\\
&&\hspace{-12em}\times\bar{J}_0(\ell\theta_j)
\left(\hat{C}_{\ell}^{yy}\hat{C}_{\ell,(ab,a'b')}^{\rm
 hh}+\hat{C}_{\ell,(ab)}^{y{\rm h}}\hat{C}_{\ell,(a'b')}^{y{\rm h}}\right),\label{eqn:xicov}
\end{eqnarray}
where $\bar{J}_0(\ell\theta_i)$ is the average of $J_0(\ell\theta)$ over
the annulus around $\theta_i$. However, due to the exponential form of the noise
term for $\hat{C}_{\ell}^{yy}$, Eq.~(\ref{eqn:xicov}) leads to divergent
results. To resolve this apparent problem, 
we instead calculate the covariance for
the stacked beam-smoothed SZ
profile, which is actually a direct observable. 
To calculate the beam-smoothed $y$-profile, 
we only need to multiply the integrand of Eq.~(\ref{eqn:xiyh}) by
$B_{\ell}$ to account for the beam convolution, and to get the covariance, we only need to multiply that of
Eq.~(\ref{eqn:xicov}) by $B_{\ell}^2$, which removes the exponential
divergence.
Note that the covariance given by Eq.~(\ref{eqn:xicov}) does
not vanish when $\theta_i\ne \theta_j$, i.e. the $y$-profiles at
different angular bins are correlated with one another 
even when the field is Gaussian.

\begin{figure*}[htb]
\vspace{0mm}
\resizebox{160mm}{!}{\includegraphics{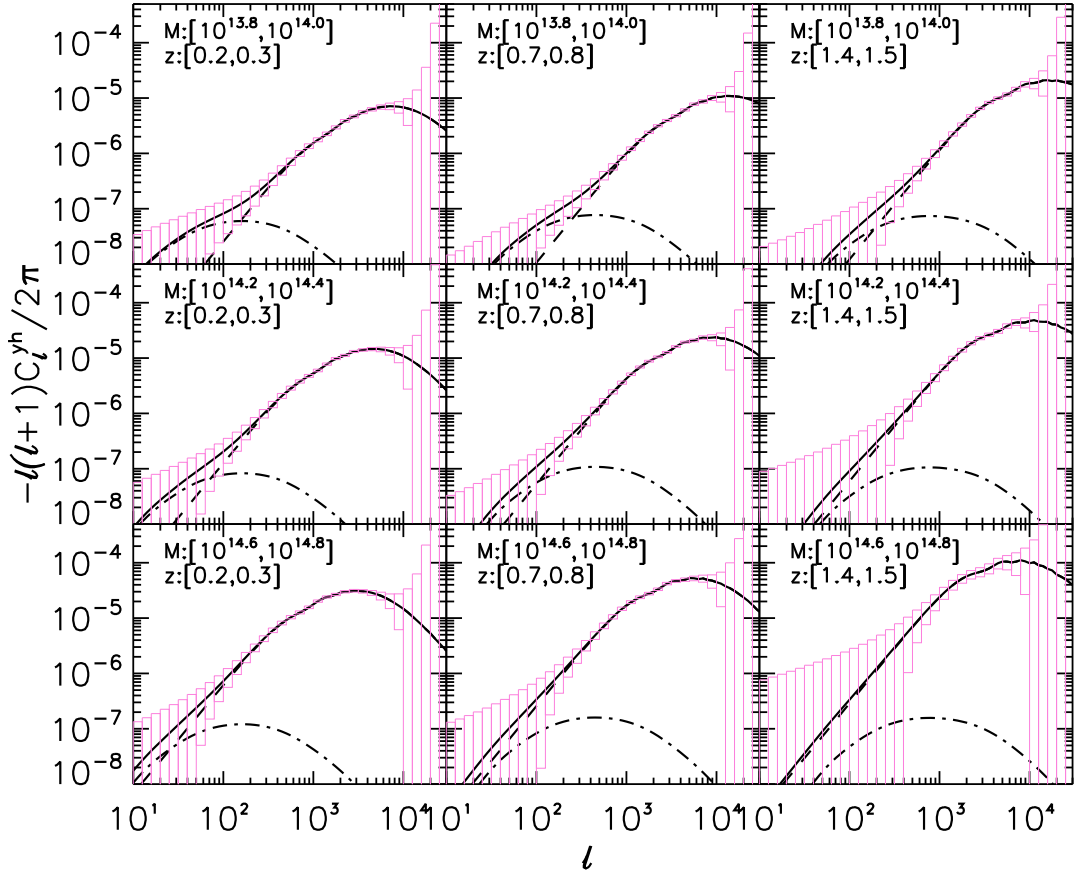}}
\vspace{0mm}
\caption{\label{fig:clyh} The cross power spectra of the SZ signal in
 the RJ band and the distribution of galaxy clusters in different
 redshift and mass bins. From left to right, the three columns are
 for clusters in the redshift bins of [0.2,0.3], [0.7,0.8] and
 [1.4,1.5], and from top to bottom, the rows are for clusters in the
 mass bins of $[10^{13.8},10^{14.0}]$, $[10^{14.2},10^{14.4}]$,
 $[10^{14.6},10^{14.8}]h^{-1}M_{\odot}$, respectively. The dashed and
 dot-dashed lines are the one-
and two-halo term contributions, while the solid lines show the total.
The boxes around each curve show $\pm 1\sigma$ statistical
 uncertainties in the power spectrum measurement expected for
a 2500 deg$^2$ SPT-DES like survey, calculated as we describe in Sec. \ref{subsec:covariance}.}
\end{figure*}

\begin{figure*}[htb]
\vspace{0mm}
\resizebox{160mm}{!}{\includegraphics{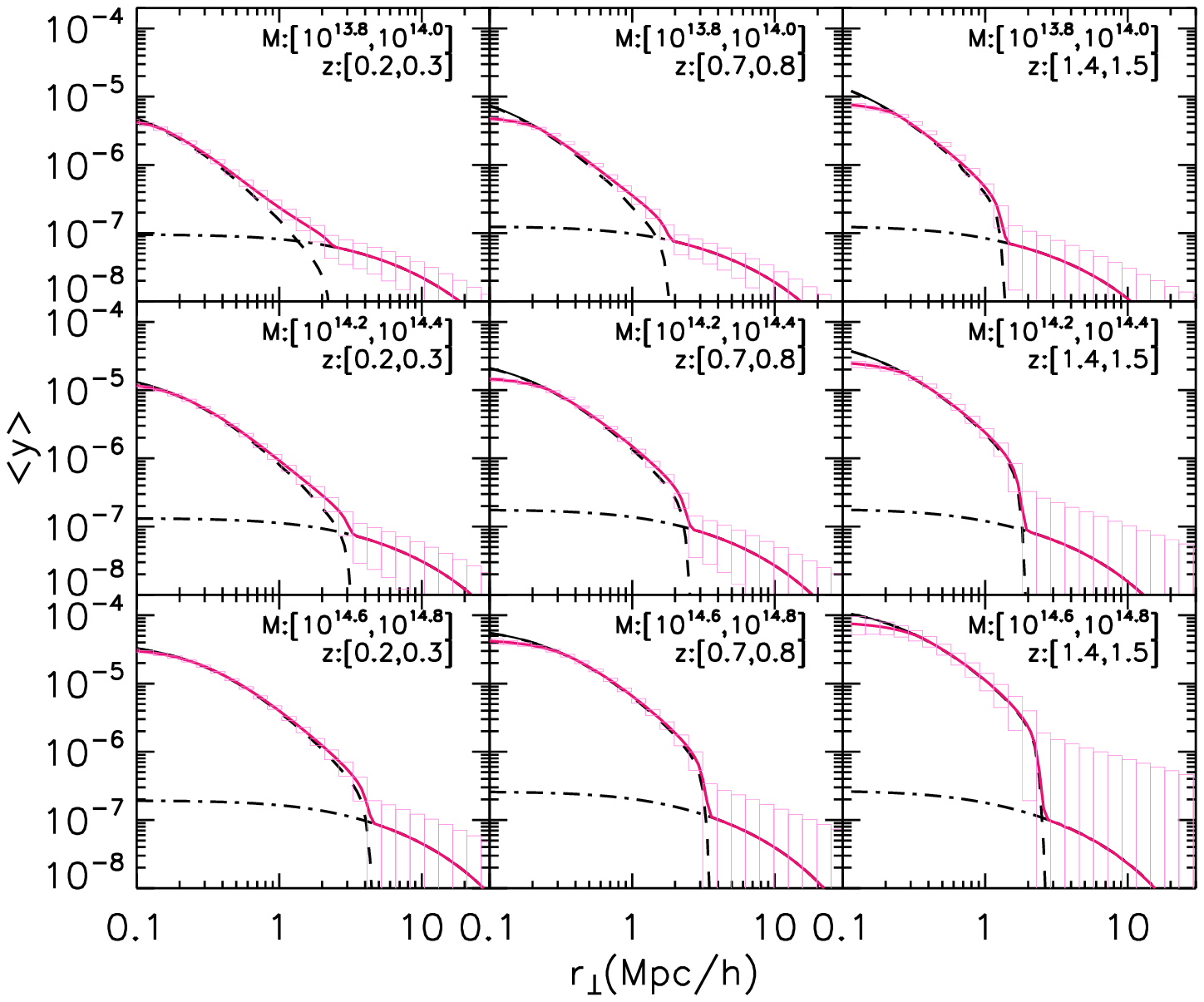}}
\vspace{0mm}
\caption{\label{fig:xiyh} Similar to Figure~{\ref{fig:clyh}}, { but here
showing the stacked $y$-profiles for galaxy clusters in different redshift
and mass bins, where the projected radius $r_{\perp}$ is converted from the
angular separation $\theta$ by using the angular diameter distance for
the mean redshift of the cluster bin. 
The error boxes are for 
the
beam-smoothed stacked profiles for the SPT-DES like survey, where we use the beam for the 150GHz SPT channel. Note that the
errors at different radii are correlated.}
  }
\end{figure*}

\section{Forecasts of the Stacked SZ Measurements}
\label{sec:detection}

In this section, we estimate and discuss the detectability of the
cross-correlation between the SZ map and the cluster distribution
from ongoing and upcoming combined CMB and optical surveys.

\subsection{Survey Parameters}

\begin{table}
\caption{\label{tab:cmb}CMB Experimental Specifications}
\begin{center}
\begin{tabular}{ c c c c c }
\hline\hline
Experiment & $\Omega_s$(deg$^2$) &$\nu$(GHz) &$\theta_{\rm FWHM}$($'$)&$\sigma_T(\mu K)$ \\
\hline
SPT\cite{wil01} & 2500 & 95 & 1.6 & 26.3\\
 & & 150 & 1.1 & 16.4\\
 & & 220 & 1.0 & 85\\
Planck\cite{Planck} & 41253 & 30 & 33 & 5.5\\
 & & 44  & 24 &7.4\\
 & & 70  & 14 &12.8\\
 & & 100 & 10 &6.8\\
 & & 143 &7.1 &6.0\\
 & & 217 &5.0 &13.1\\ 
 & & 353 &5.0 &40.1\\
 & & 545 &5.0 &400.6\\
 & & 857 &5.0 &18257.5\\
\hline
\end{tabular}
\end{center}
\end{table}

\begin{table}
\caption{\label{tab:optical}Optical Cluster Surveys}
\begin{center}
\begin{tabular}{c c c}
\hline\hline
Survey & $\Omega_s$(deg$^2$) & $z$-range\\
\hline
SDSS &  7500 & 0.1-0.3\\
DES & 5000 & 0.1-1.0 \\
LSST & 20000 & 0.1-1.4\\
\hline
\end{tabular}
\end{center}
\end{table}

\begin{table}
\caption{\label{tab:combine}Combined SZ and Optical Cluster Surveys}
\begin{center}
\begin{tabular}{c c c c}
\hline\hline
Survey & $\Omega_s$(deg$^2$) & $\ell_{\rm min}$ & $\ell_{\rm max}$ \\ 
\hline
SPT-DES & 2500 & 7 &$\sim10000$ \\ 
Planck-SDSS & 7500 & 4 & $\sim 3000$ \\ 
Planck-LSST & 20000 & 3 & $\sim3000$ \\ 
\hline
\end{tabular}
\end{center}
\end{table}

To forecast the detectability of the stacked SZ signals from ongoing and upcoming surveys, we need to
specify the survey parameters.
In
this paper, we consider the case that an optical survey, overlapped with a CMB
survey, is used to construct a catalog of clusters.
Such joint CMB and
optical surveys that we are interested in include, e.g., DES and SPT, Subaru HSC
and ACT, and SDSS or LSST and Planck.

As for the CMB surveys, specifications of each CMB experiment
considered in this paper are given in Table~\ref{tab:cmb}.
Table~\ref{tab:optical} gives the survey parameters for the optical
surveys.  Although an optical survey alone suffers from incompleteness
of cluster finding, which may lead to a complicated selection function
\cite{Koesteretal:07,Cohnetal:07}, we here assume that the optical survey can find
all the clusters in the specified ranges of mass and redshift. Obviously this is
too optimistic, and one should keep in mind that the following results
give the best available case for the stacked SZ measurements. For a more realistic study,
the selection function for a given survey needs to be properly
taken into account.
Table~\ref{tab:combine} gives the parameters for the joint CMB and optical surveys, including the overlapping area, the
minimum and maximum multipoles probed.

\subsection{Forecasts}

In  
Figures~\ref{fig:clyh} and \ref{fig:xiyh}, 
we show the stacked SZ power spectra, scaled to the RJ limit, and 
the stacked $y$-profiles 
for clusters in different redshift and mass bins.
The dashed and dot-dashed curves in each panel
are the one-
and two-halo term contributions, respectively,
while the solid curve represents the total. As can be seen, in Fourier
space, the one-halo term dominates on small scales, while the two-halo
term becomes more important on large scales, especially for the bins of
clusters with lower masses and lower redshifts. Similar results hold in real space, 
except that the one-halo term is clearly truncated at 
about three times 
the virial radius of the most massive halo in
each bin, 
as our gas profile is truncated at $r= 3 r_{\rm vir}$ (see 
\S~\ref{sec:szeffect}). Note that we have converted the angular separation $\theta$ to the projected distance separation {$r_{\perp}$} by using the (physical) angular diameter distance at the mean redshift of the bin.
Thus, for a thin redshift slice, the stacking
method probes the average projected pressure profile for the clusters stacked. 
In a statistically isotropic universe, the average 3D pressure profile
is considered to be spherically symmetric, and therefore one can convert 
the projected pressure profile to obtain the 3D profile
by using the Abel's
integral \cite[see e.g. Eq. (1) in][]{Johnstonetal:07}.

The self-similar model predicts that the amplitude of the Compton
$y$-parameter profile $y_0$ scales with the cluster's virial density
$\rho_{\rm vir}$ and virial mass $M_{\rm vir}$ as $y_0 \propto \rho_{\rm
vir}M_{\rm vir}$ (see the Appendix). 
Since $\rho_{\rm vir}$ increases with $z$, $y_0$
increases with $z$ as well, in addition to its increasing with $M_{\rm
vir}$. These trends agree with 
what we see in Figures~\ref{fig:clyh} and \ref{fig:xiyh}: the one-halo term for
the cross correlation gets greater for cluster bins with higher $z$ and
larger $M$, though the gas model we use is not exactly self-similar
(broken by the dependence of the concentration parameter $c$ on $M_{\rm
vir}$ and $z$). As the clusters become more massive and their redshifts get smaller, 
the angles they extend become larger, hence in real space, the one-halo term terminates at a larger angular scale, while in Fourier space, it peaks at a smaller $\ell$. 

The error boxes around each curve in Figures~\ref{fig:clyh} and
\ref{fig:xiyh} show the $1\sigma$ statistical uncertainties for
measuring the stacked SZ signals at each multipole (radial) bin. These are expected for a 2500 deg$^2$ SPT-DES like survey. Note the error boxes at different radial bins for the stacked $y$-profile are
correlated. These figures show that the stacking method can lead to a significant detection of the SZ signal over a wide range of
multipoles or radii. In particular, it
allows for exploration of the SZ signal at large radius, even around or beyond $r_{\rm vir}$, as can be seen from Figure~\ref{fig:xiyh}. 
We find the variance of $C_{\ell}^{y{\rm h}}$ (Eq.~\ref{eq:cly_cov}) is mostly dominated by the term
$\hat{C}^{\rm hh}_{\ell}\hat{C}^{yy}_{\ell}$. At $\ell \gsim 100$, it comes mainly from
$
\hat{C}^{yy}_{\ell}/\bar{n}^{2D}_{ab}$, and at even smaller scales such as $l\gsim 10^{4}$,
instrumental noise dominates the error budget.

To be more quantitative, 
we calculate the cumulative signal-to-noise square
$(S/N)^2$ for measuring
the stacked SZ signals for the clusters binned by their redshifts and masses,
which in Fourier space is an integration over a given range of multipoles, defined as
\begin{equation}
\left(\frac{S}{N}\right)^2=\sum_{\ell}C_{\ell,(ab)}^{y\rm{h}}{\rm
 Cov}^{-1}\left(C_{\ell,(ab)}^{y{\rm h}},C_{\ell,(ab)}^{y{\rm
	   h}}\right)C_{\ell,(ab)}^{y{\rm h}}.
\label{eq:sn}
\end{equation}
The $(S/N)^2$ is equivalent to the information content available from a Fisher
matrix analysis~\cite{TakBhu09,TegTay97}: given a template for the
interested power spectrum, it gives the significance that the amplitude
of the power spectrum is detected to be non-zero. We notice that the $(S/N)^2$ is proportional to the survey area
$\Omega_{\rm s}$ (assuming a simple survey geometry).

\begin{figure}[htb]
\resizebox{90mm}{!}{\includegraphics{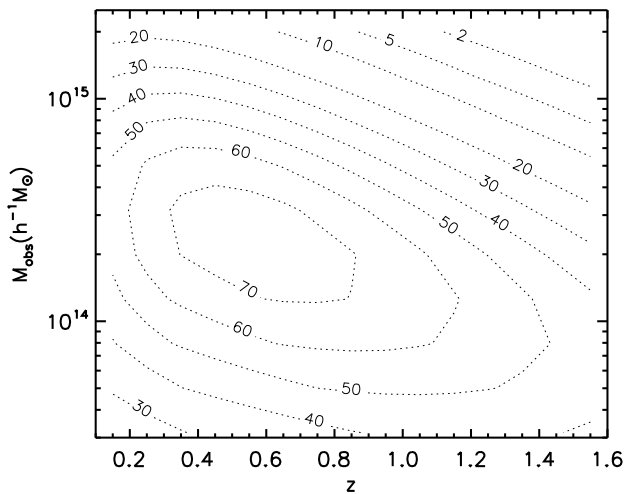}}
\caption{\label{fig:sptdes} Contours of the total signal-to-noise ratio (S/N) expected for measuring the cross-correlation between the SZ effect and the
 distribution of galaxy clusters assuming a SPT-DES like survey (see
 Eq.~\ref{eq:sn} for the definition of the S/N). 
The clusters are binned in redshift with $\Delta z=0.1$ and in mass with
 $\Delta \log (M_{\rm obs})=0.2$, respectively. }
\label{fig:sn_spt}
\end{figure}

\begin{figure}[htb]
\resizebox{90mm}{!}{\includegraphics{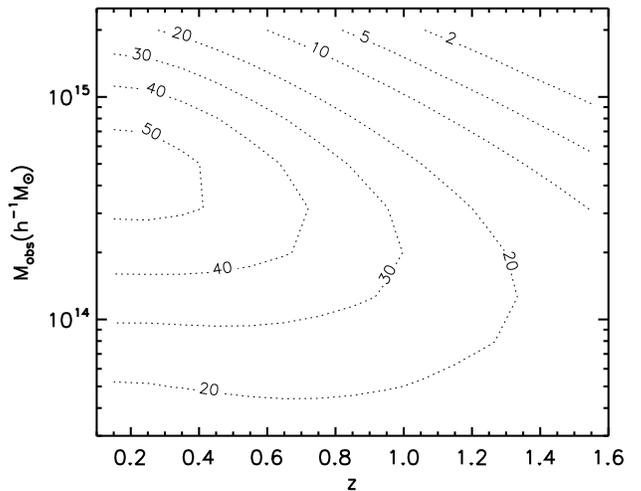}}
\caption{\label{fig:plancksdss} Similar to the previous plot, but for 
a Planck-SDSS like survey. Note here the contours are displayed out to much higher redshift regions than those available from the SDSS, which is z:[0.1,0.3].
}
\end{figure}

\begin{figure}[htb]
\resizebox{90mm}{!}{\includegraphics{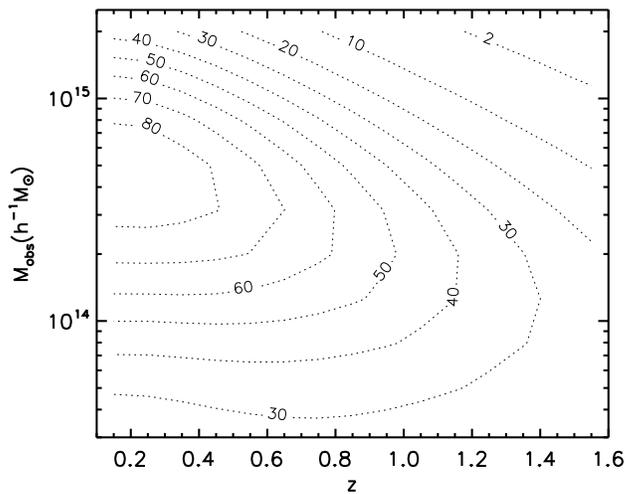}}
\caption{\label{fig:plancklsst}Similar to Figure~\ref{fig:sptdes}, but 
for a Planck-LSST like survey.
}
\end{figure}

In Figures~\ref{fig:sptdes}-\ref{fig:plancklsst}, we plot the contours of
the $S/N$ for the SPT-DES, Planck-SDSS, and Planck-LSST like surveys, respectively. 
Again note that we have assumed all the clusters in a given mass
and redshift bin can be found from the survey, i.e. 100\% completeness and
efficiency of cluster finding.
As we noticed before, the amplitude of the cross correlation increases when
the clusters in the bin have larger masses or higher redshifts. At the
same time, the number of clusters in the bin decreases with their masses,
and also decreases with their redshifts, though it increases with the
redshift at first. While the $S/N$ increases when the signal is higher,
it decreases when the shot noise is larger. The result is that the $S/N$
peaks for some cluster bin at intermediate redshift and intermediate
mass -- the sweet spot, as can be seen from these three figures. From SPT to
Planck, the resolution of the instruments gets worse, and the sweet spot
shifts to lower redshift and higher mass, where the clusters extend
larger angles such that their profiles can be better resolved. 
In
particular, it would be interesting to notice that the
arcminute-resolution CMB experiment such as SPT has the sweet spot
at around $z\simeq 0.6$ and $M\simeq 2\times 10^{14}h^{-1}M_\odot$. 
From SDSS
to LSST, the $S/N$ increases due to the larger survey area. Note $S/N \propto \sqrt{f_{\rm sky}}$, provided the probed $\ell$ ranges
are similar.

Recently, the Planck team has shown their high significance detection of the SZ signals for the SDSS maxBCG clusters~\cite{pl3}, down to the low mass systems. Their analysis is different from what we are doing in this paper. Instead of stacking the SZ maps around the center of each cluster, they derived the integrated SZ signal $Y_{500}$ for each cluster by employing a multi-frequency matched filter, and then calculated the weighted mean of the signal. The $S/N$ of their detection ranges roughly 4$\sim$15 for the cluster bins that are relatively narrower in mass and wider in redshift than ours. Despite the differences in analysis, their results are less optimal than ours. This may be because: the Planck results are from its first-year survey, so the instrumental noises are larger than their expected values, which are what we adopt in this paper; our results neglect the impurity and incompleteness of the maxBCG cluster catalog, hence tend to be optimal. Interestingly, we notice the $S/N$ of the Planck's results also peaks at some intermediate mass $\sim 2\times 10^{14}h^{-1}M_{\odot}$ (corresponding to $N_{200}\sim30$), which is smaller than what we find by roughly a factor of $2$.

\subsection{Residual Noise Contamination}
\label{sec:discuss}

Our estimation for the noise in the extracted SZ signal so far includes only the instrumental
noise. This is optimistic, since there are various sources of contamination to
the SZ effect, e.g. the point sources (both the Poisson distributed and
clustered components)~\cite{Lue10,Hal10,Fowleretal:10,PlanckCIB:11}, and the primary
CMB fluctuations, which may leave some residual components even after
the SZ signal has been maximally extracted. In this subsection, we consider
two types of residual noise: those from the point
sources and from the primary CMB, and see how the $S/N$ of the stacked
SZ signal is degraded.

In the following, we give a simple estimate on the impact of
these two types of residual noise on the measurement of the stacked SZ signal.
The primary CMB fluctuations do not correlate with the
distribution of clusters. Therefore, we assume the residual CMB does not contaminate the stacked SZ signal, but adds to its noise. We also assume the residual CMB has the same power spectrum as the primary CMB, for simplicity. As for the contamination from point sources, we
need to consider two populations: radio sources such as galaxies hosting
active galactic nuclei (AGNs), and infrared sources such as the dusty
star-forming galaxies. Although radio sources tend to reside in clusters, their contamination to the SZ signal 
is unlikely to be significant (10\% at most for low-$z$
clusters) \cite{Linetal:09}. On the other
hand, infrared sources dominate the point source contamination
to the SZ power, with both a Poisson-distributed and a clustered component. However, it is likely that the majority of contamination
comes from galaxies at high redshifts such as $z\gsim
2$ \cite{PlanckCIB:11}, which do not correlate with the
massive halos at $z\lsim 1.5$ as are considered in this paper. 
Hence we
assume that the residuals from point sources do not contaminate
$\hat{C}_\ell^{yh}$, but add to $\hat{C}_\ell^{yy}$. Specifically, we assume the residuals are Poisson distributed, and estimate their impact by simply doubling the instrumental noise.

\begin{figure}[htb]
\resizebox{90mm}{!}{\includegraphics{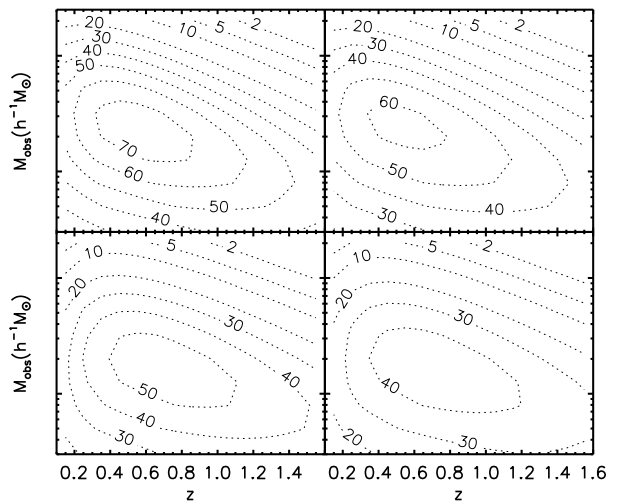}}
\caption{\label{fig:sptdesrp} 
Similar to Figure~\ref{fig:sptdes}, but using different models for
the noise in the SZ field. 
The upper-left panel is the same as Figure~\ref{fig:sptdes}, shown here for comparison, which includes only the instrumental noise, the
 upper-right has the residual noise from point
 sources added in, the lower-left has that from the primary CMB added
 in, while the lower right has both (see text for the details).}
\end{figure}

The
results of our calculation are shown in Figure~\ref{fig:sptdesrp}. These
are the $S/N$ contours for a SPT-DES like survey. In the upper right
panel, we include the residual noise from the point sources, in the
bottom left, we include that from the primary CMB, while in the bottom
right, we include both. To easily see the effects of these residual
noises, we show in the upper left panel our results before, i.e., with
instrumental noise only. As is expected, the $S/N$ is reduced when
either of the residual noises is included. According to our calculation for the residual power spectra, the primary CMB causes a larger reduction in the $S/N$ than the point sources, which perhaps indicates the necessity to remove the CMB. From Figure~\ref{fig:clyh}, we can also see, on scales where the primary CMB dominates ($\ell \lsim 3000$), there is non-negligible contribution to the total $S/N$. Finally, we find, even in the most pessimistic case we study, the $S/N$ for the detection of the stacked SZ signals for the less massive clusters fortunately remains substantial ($\gsim 20$).

\section{Discussion}
\label{sec:discuss}

In this paper, we have developed the halo model based method to 
compute the stacked SZ profile of the galaxy clusters, measured by cross-correlating the SZ
map and the clusters' distribution, and the method to estimate the expected
signal-to-noise ($S/N$) ratio for its detection from a given survey. Such stacked SZ signals can be obtained,
e.g., by combining the arcminute-resolution CMB experiments with optical
surveys such as the combination of the DES and SPT, or the Subaru HSC and ACT surveys.

Although the measured CMB signal is by nature two-dimensional,
cross-correlating the CMB map with clusters of known redshifts allows
for a {\em tomographic} reconstruction of the SZ signals as a function of redshift. In addition, simultaneous reconstruction of the SZ signals as a function of cluster mass is also allowed, if the clusters' masses are also known. This would be very useful for carrying out the precision cosmology
with cluster based experiments as well as studying the cluster astrophysics, in particular, its time evolution \citep[see, e.g.,][]{FanHai07}. 

The stacked SZ signals will be very complementary to the stacked
lensing signals \cite{Johnstonetal:07,Okabeetal:10}, if both methods
study the same catalog of clusters.  One of the crucial sources of systematic in cluster cosmology is from the mass-observable relation. To attain
the precision of cluster cosmology, we need a well-calibrated mass-observable
relation. Combining the stacked SZ and lensing signals for the same
catalog of clusters will provide a powerful means of 
studying the {\em mean} relation between cluster mass and SZ signal as a function of redshift and mass (more precisely, the optical richness) from the observation. As we have shown, the stacking method can
significantly reduce the statistical errors, and allows one to probe the
SZ or lensing signals for the low mass clusters or to large angular scales \cite[also see][]{OguTak11}. However,
the scatters around the mean mass-observable relation still need to be
understood, and detailed studies of the target cluster sample combining
various observables will also be needed \cite[e.g., see][for such a
multi-wavelength study]{Marroneetal:11}.

Combining the stacked SZ and lensing signals will be powerful
for studies of cluster physics as a function of cluster mass and
redshift. Due to the statistical symmetry of the
universe, the stacked pressure and mass profiles can be considered to be
spherically symmetric in an average sense. Hence it would be of great interest to construct the cluster's baryon
fraction profile observationally, and then address at
which radii the cosmic mean baryon fraction is reached. Since the
intra-cluster gas may be blown off to outer radii 
by various feedback effects from e.g.
AGN or supernova, which involve complicated astrophysical
processes, such a model-independent method of studying the baryon
fraction would be very useful \citep[see][for the related
discussion]{Afshordietal:07,Umetsuetal:09}.

The SZ signals from galaxy clusters would be smaller if there is a significant fraction of non-thermal pressure support from random gas motions, whose existence could help to
explain the apparent conflict between the amplitude of the observed SZ
power spectrum with theoretical prediction \cite{Shawetal:10}. Such 
non-thermal pressure support is expected to be more significant at the outskirts for given mass-scale halos \cite{Lau09}. Again the stacking
method is very powerful for studying the SZ signals at such outer radii, which
otherwise would be very difficult due to the faint signals.

Throughout this paper, we have been using the relatively simple analytical model from \cite{KomSel02} for the cluster
pressure profile, given as a function of halo mass and redshift. 
It is known this model overestimates the measured SZ power spectrum amplitude (by about a factor of 2), which however would not alter our conclusion of a significant S/N for the stacked SZ measurement that is evident from Figure~3-5. 
However, a
more careful study will be definitely needed, e.g., by using more realistic pressure profiles \citep[see, e.g.,][]{Arn10}, or hydrodynamic
simulations, in order to realize the genuine power of the stacking
method. 

\begin{acknowledgments}
We thank August Evrard, Eiichiro Komatsu, Erwin Lau, and Jeff McMahon for useful discussions, and Tomasz Biesiadzinski, Brian Nord for comparing our work with their preliminary numerical studies. W.F. is supported by the NSF under contract AST-0807564, and by the NASA under contract NNX09AC89G.
 This work is supported in part by the Michigan Center for Theoretical Physics, the Grants-in-Aid for Scientific
Research Fund (No. 23340061), JSPS Core-to-Core Program
``International Research Network for Dark Energy'', World Premier
International Research Center Initiative (WPI Initiative), MEXT, Japan,
and the FIRST program ``Subaru Measurements of Images and Redshifts
(SuMIRe)'', CSTP, Japan.
\end{acknowledgments}

\vfill
\bibliographystyle{physrev}
\bibliography{stack}

\onecolumngrid
\appendix
\begin{center}
  {\bf APPENDIX}
\end{center}

In the self-similar model, the gas profiles of all the clusters are considered identical with appropriate choice of normalization factors that are the powers of the global quantities e.g. mass $M_{\rm vir}$, radius $r_{\rm vir}$. For example, the temperature profile can be expressed as $T(r)/T_{\rm vir}\propto f_{T}(r/r_{\rm vir})$, with $T_{\rm vir} \propto M_{\rm vir}/r_{\rm vir}$(from the virial theorem) and $f_{T}(r/r_{\rm vir})$ a universal function. Similarly, the pressure profile can be expressed as $P(r)/P_{\rm vir}\propto f_{P}(r/r_{\rm vir})$, with $P_{\rm vir}\propto \rho_{\rm vir}T_{\rm vir}$ and $f_{P}(r/r_{\rm vir})$ a universal function. Hence, the $y$-profile is given by
\begin{eqnarray}
y(\theta)&&=\frac{\sigma_T}{m_e c^2}\int_{-\sqrt{r_{\rm vir}^2-r_{\perp}^2}}^{\sqrt{r_{\rm vir}^2-r_{\perp}^2}} P\left(\sqrt{\ell^2+r_{\perp}^2}\right)d\ell, \nonumber \\ &&\propto P_{\rm vir}r_{\rm vir}f_y(r_{\perp}/r_{\rm vir}),\nonumber \\ &&\propto \rho_{\rm vir}M_{\rm vir}f_y(\theta/\theta_{\rm vir}),
\end{eqnarray}  
where $\theta_{\rm vir}=r_{\rm vir}/d_A$, with $d_A$ the angular diameter distance, and $f_y$ is a universal function. As can be seen, the amplitude of the self-similar $y$-profile is proportional to $\rho_{\rm vir}M_{\rm vir}$.
\end{document}